\documentclass[twocolumn,
english,
aps,
pra,
10pt,
longbibliography,
superscriptaddress,
amsmath,
amssymb,
floatfix
]{revtex4-2}
\usepackage[utf8]{inputenc}
\usepackage{microtype}
\usepackage{braket}
\usepackage{tabularx}
\usepackage{multirow}
\usepackage{physics}
\usepackage{graphicx}
\usepackage{lipsum}
\usepackage{dsfont}
\usepackage{mathtools}
\usepackage[colorlinks=true,urlcolor=blue,citecolor=blue,linkcolor=blue]{hyperref}
\usepackage{natbib}
\usepackage{cleveref}
\usepackage{sidecap}
\usepackage{makecell}
\usepackage[linesnumbered,lined,commentsnumbered]{algorithm2e}
\Crefname{algocf}{Algorithm}{Algorithms}
\usepackage{layouts}
\usepackage{dcolumn}
\usepackage{bm}
\usepackage[normalem]{ulem}

\begin{document}
\title{Learning How to Dynamically Decouple}
\date{\today}

\author{Arefur Rahman}
\affiliation{School of Electrical, Computer, and Energy Engineering, Arizona State University, Tempe, Arizona 85287, USA}

\author{Daniel J. Egger}
\affiliation{IBM Quantum, IBM Research Europe - Zurich, R\"uschlikon 8803, Switzerland}

\author{Christian Arenz}\email{carenz1@asu.edu}
\affiliation{School of Electrical, Computer, and Energy Engineering, Arizona State University, Tempe, Arizona 85287, USA}

\begin{abstract}
Current quantum computers suffer from noise that stems from interactions between the quantum system that constitutes the quantum device and its environment. 
These interactions can be suppressed through dynamical decoupling to reduce computational errors.
However, the performance of dynamical decoupling depends on the type of the system-environment interactions that are present, which often lack an accurate model in quantum devices. 
We show that the performance of dynamical decoupling can be improved by optimizing its rotational gates to tailor them to the quantum hardware. 
We find that compared to canonical decoupling sequences, such as CPMG, XY4, and UR6, the optimized dynamical decoupling sequences yield the best performance in suppressing
noise in superconducting qubits. 
Our work thus enhances existing error suppression methods which helps increase circuit depth and result quality on noisy hardware.
\end{abstract}

\maketitle

\section{Introduction\label{sec:intro}}
Noise currently limits quantum computers from harnessing their full potential.  
In the long term, quantum error correction is expected to overcome this issue~\cite{bravyi2024qec}.
In the short term, noise mitigation and suppression techniques are critical to improve quantum device performance~\cite{cai2023em}. 
Error mitigation is designed to reduce noise, typically, in expectation values~\cite{cai2023em, Temme2017}.
The computation is executed on the noisy quantum computer multiple times to either extrapolate to a zero-noise limit~\cite{kim2023b}, or to cancel the noise on average~\cite{vandenBerg2023}.
By contrast, error suppression methods, such as dynamical decoupling~\cite{viola1998dynamical} and pulse-efficient transpilation~\cite{earnest2023pulse}, reduce the presence of noise directly in the quantum circuits.

In this work we focus on the well-established noise and error suppression technique  dynamical decoupling (DD).
Inspired by nuclear magnetic resonance spectroscopy (NMR) \cite{c2_haeberlen1968coherent}, the theory of DD was first developed by Lorenza Viola, Emanuel Knill, and Seth Lloyd in 1998~\cite{viola1998dynamical} as an open-loop control technique. 
DD suppresses errors by decoupling the system from its environment through the application of a sequence of pulses which, in the ideal case, compose to the identity.
The utility of DD has been demonstrated in a wide range of quantum systems, such as coupled nuclear and electron spins~\cite{morton2006}, trapped ions~\cite{biercuk2009optimized}, electron spins~\cite{du2009preserving}, and superconducting qubits~\cite{ezzell2022dynamical}. Since the introduction of the DD framework in the 90's, DD has also become a viable method to suppress noise and errors in quantum computing~\cite{lidar2013quantum}.  
For example, DD can suppress crosstalk~\cite{tripathi2022suppression, seif2024suppressing} and improve the performance of superconducting qubit based quantum devices~\cite{pokharel2018, ezzell2022dynamical} in general.  
DD sequences can be designed from first principles~\cite{uhrig2007} or with numerical simulations, leveraging tools such as genetic algorithms~\cite{quiroz2013} and machine learning~\cite{august2017}. 

\begin{figure}[!h]
     \centering    \includegraphics[width=\columnwidth]{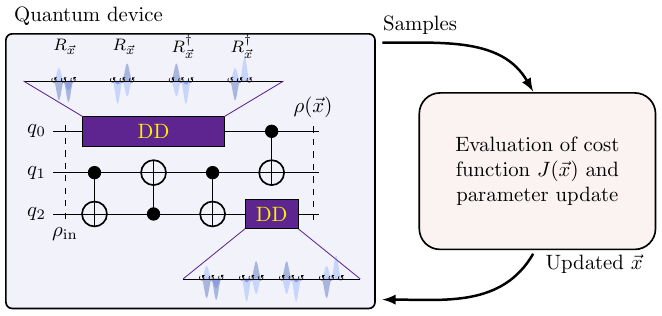}
     \caption{\textbf{Illustration of LDD}.
     The left and the right block indicate a quantum device and a classical computer, respectively, working in tandem. 
     A DD sequence parameterized by $\vec{x}$ is added to idling qubits in a given quantum circuit, here exemplified by a SWAP gate on $(q_1, q_2)$ followed by a CNOT on $(q_0, q_1)$.
     The arbitrary single-qubit rotations $R_{\vec{x}}$ in the DD sequence are realized by three parameterized virtual $Z$-gates (circular arrows) and two $\sqrt{X}$ gates (pulse with in-phase and quadrature shown in light and dark blue, respectively).
     The noisy output state $\rho(\Vec{x})$ is compared to the expected ideal state $\rho_\text{ideal}$ through the cost function $J(\vec{x})$. 
     A classical optimization algorithm minimizes $J(\Vec{x})$ to find the parameters of the gates in the DD sequence that best fit the circuit.}
    \label{fig1:dd_workflow}
\end{figure}

In recent years, DD has become a major error suppression method for noisy superconducting quantum computers~\cite{pokharel2018}.
Indeed, the method is easy to apply as a simple transpilation pass that inserts delays and pulses into a quantum circuit. 
Furthermore, simple sequences such as $X-X$ already yield excellent results~\cite{jurcevic2021qv}.
More elaborate sequences, such as  staggered $X-X$~\cite{Shirizly2024} and staggered $XY4$, improve, for instance, the execution of dynamic circuits by cancelling cross-talk~\cite{vazquez2024scaling}. Here $X$ and $Y$ are $\pi$-rotations around the qubit's $x$ and $y$-axis, respectively.
Crucially, the performance of a DD sequence depends on the interactions present in the quantum hardware.
In superconducting qubits~\cite{krantz2019}, a good model of these interactions is typically not known, a case familiar to optimal control~\cite{egger2014adaptive, PhysRevA.102.062605, PhysRevA.101.032313, magann2021} that can be overcome with closed-loop optimization~\cite{kelly2014oct, werninghaus2021}. Similarly, it is possible to tailor DD sequences to the quantum hardware at hand by learning them through genetic algorithms~\cite{tong2024}. However, in hardware, DD pulses are not perfect, potentially introducing additional noise and errors through the controls that implement the DD pulses, thereby diminishing the quality of the designed DD sequence. While control noise-robust DD sequences exist \cite{lidar2013quantum}, e.g., suppressing errors in rotational angles of the DD gates, such sequences typically require knowledge of the type of control error that is introduced.

To overcome these limitations, in this work we tailor the DD sequences to the hardware and quantum circuits to execute. The goal is to design DD sequences that suppress potentially unknown pulse imperfections and noise.
This is achieved by optimizing the rotational angles of the gates in the DD sequence in a closed-loop with the quantum hardware.
A classical optimizer is fed the cost function value that is reconstructed from quantum samples  and that is sensitive to the quality of the DD pulses, see Fig.~\ref{fig1:dd_workflow}.

This manuscript is structured as follows.
We introduce in Sec.~\ref{sec:dd} hree commonly employed DD sequences CPMG, XY4 and UR6.
Next, in Sec.~\ref{sec:ldd} we develop the theoretical framework of how optimal parameters in DD sequences are found on quantum hardware, which we refer to as learning dynamical decoupling (LDD).
We demonstrate in Sec.~\ref{sec:casestuddies} the utility of LDD on IBM Quantum hardware by comparing the performance of LDD to CPMG, XY4, and UR6 to suppress noise in two experiments.
We show that LDD outperforms CPMG, XY4, and UR6 for suppressing noise present during mid-circuit measurements and noise resulting from increasing the depth of a quantum circuit. 
We conclude in Sec.~\ref{sec:conclusion}.

\section{Background: Dynamical Decoupling\label{sec:dd}}
DD is a well known strategy that reduces noise by suppressing unwanted interactions with the environment or undesired couplings of the controls with the system. For an overview of DD strategies we refer to \cite{lidar2013quantum}.
DD can in general suppress generic interactions through pulses that rotate around multiple axis~\cite{c14_viola1999dynamical}.
However, DD is most resource efficient when tailored to the specific type of interactions at hand~\cite{biercuk2009experimental, berglund2000quantum}.
Furthermore, the most effective DD sequence depends on the noise type present in the physical system. 

\begin{figure}[!ht]
     \centering    \includegraphics[width=0.9\columnwidth]{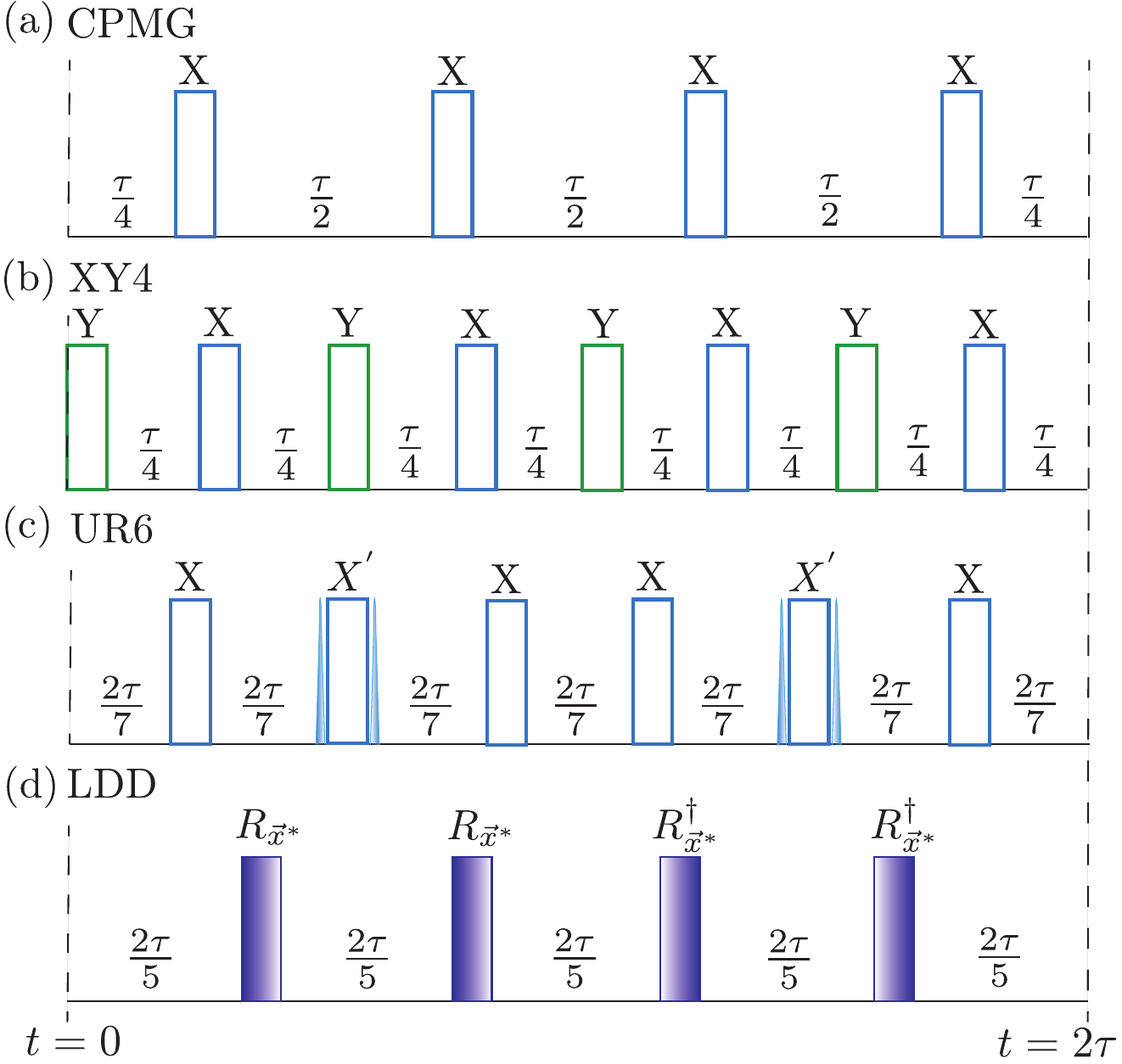}    
     \caption{\textbf{Schematic diagram of DD sequences}. 
     (a) Two concatenated CPMG sequences. (b) Two concatenated XY4 sequences. 
     (c) One UR6 sequence. (d) Two concatenated LDD sequences.
     All sequences are depicted for an idle period of $2\tau$ while the gate implementation time is not shown. 
     When running on hardware, the idle times are reduced to account for the duration of the physical pulses of the DD gate. 
     CPMG uses only $X$ gates (blue rectangle) whereas XY4 uses both $X$ and $Y$ gates (green rectangles). Here $X$ and $Y$ are $\pi$-rotations around the qubits $x$-axis and $y$-axis, respectively.
     UR6 uses $X^{'}$ gates that consists of two phase gates with angles $2\pi/3$ (shown as blue pulses) with opposite rotational directions that are inserted before and after an $X$ gate, see Eq. \eqref{eq2_2:UR6}.  
     LDD uses parameterized gates, $R_{\Vec{x}^*}$ (purple rectangles) where the optimal rotation angles $\Vec{x}^*$ are found by classical optimization routines performed in tandem with a quantum device.
     }
    \label{fig2:dd_sequence}
\end{figure} 

The spin echo \cite{hahn1950spin} in NMR can be seen as a DD experiment where a single Pauli $X$ gate refocuses coherent errors. The ``Carr–Purcell–Meiboom–Gill'' (CPMG) DD sequence is an extension of the spin echo with two symmetric insertions of an $X$ gate~\cite{carr1954effects, meiboom1958modified}.
The resulting pulse sequence is
\begin{equation} \label{eq1:CPMG}
\text{CPMG} \equiv \tau/2 - X - \tau - X - \tau/2. 
\end{equation}
Here, the total duration of the free evolution is $2\tau$ and $- \tau -$ indicates an idle period of duration $\tau$.
Multiple CPMG sequences can be concatenated one after another, see Fig.~\ref{fig2:dd_sequence}(a).
While the CPMG sequence can suppress homogeneous dephasing along one axis, it cannot suppress noise stemming from generic system-environment interactions. 
By contrast, the XY4 DD sequence~\cite{maudsley1986modified}, defined by 
\begin{equation} \label{eq2:XY4}
\text{XY}4 \equiv Y - \tau/2 - X - \tau/2 - Y - \tau/2 - X - \tau/2,
\end{equation}
is a universal DD sequence that can suppress generic system-environment interactions.
Two concatenated XY4 sequences are shown in Fig.~\ref{fig2:dd_sequence}(b). Both CPMG and XY4 are sensitive to pulse imperfections~\cite{ali2013robustness}. 
Such errors can be suppressed with a universal rephasing (UR) DD sequence by inserting properly tailored phase gates into, e.g., a CPMG to obtain the symmetric UR6 DD sequence~\cite{Genov2017}
\begin{align} \label{eq2_2:UR6}
\begin{split}
\text{UR}6 \equiv {}& 2\tau/7 - X - 2\tau/7 - X^{'} - 2\tau/7 - X - 2\tau/7 \\
&  - X - 2\tau/7 - X^{'} - 2\tau/7 - X - 2\tau/7.\\
\end{split}
\end{align}
The UR6 sequence has four $X$ gates and two phased $X$ gates denoted by $X^{'} = R_{z}\left(-\frac{2\pi}{3}\right)XR_{z}\left(\frac{2\pi}{3}\right)$ where $R_{z}(\theta)$ is a rotation around the qubits $z$-axis by an angle $\theta$.

DD sequences with higher order protection can be built with additional single-qubit pulses, for example, by concatenating existing sequences~\cite{khodjasteh2005} or considering semi-classical noise models~\cite{Genov2017}.
However, the effectiveness of DD sequences critically depends on the type of noise, i.e., resulting from the detrimental interactions that are present in the system, which is often challenging to infer in superconducting quantum devices~\cite{krantz2019}. 
For example, Ref.~\cite{ezzell2022dynamical} shows that the performance of different DD sequences is device dependent even though the devices all leverage the cross-resonance interaction~\cite{Sheldon2016}.
Furthermore, pulse imperfections, such as a detuning in frequency or amplitude, degrade the performance of a DD sequence~\cite{Genov2017}.
As such, the performance of a DD sequence can vary substantially across different devices and even time, if hardware parameters drift.

\section{Learning how to suppress noise\label{sec:ldd}}
We use tools from closed-loop optimal control \cite{egger2014adaptive} and optimization \cite{kelly2014oct, das2021adapt,  werninghaus2021,tong2024} to learn optimal DD sequences without precise knowledge of a noise model. 
Tong \emph{et al.}~\cite{tong2024} demonstrate the usefulness of this approach by optimizing with a genetic algorithm the placement of DD gates on idling qubits in a quantum circuit to achieve, for instance, a higher success probability in the Bernstein-Vazirani algorithm compared to canonical DD sequences.
In their approach, they chose DD gates from the fixed set $\{I_\pm, X_\pm, Y_\pm, Z_\pm\}$ where $I_+$, $X_+$, $Y_+$, $Z_+$ are the Pauli matrices and the minus sub-script indicates an added phase of $\pi$.

DD gates themselves are prone to errors, such as errors in the rotation angles.
Furthermore, the optimal rotation axis of the gates in the DD sequence may depend on the type of noise in the system.  
We address these issues by adopting a different optimization approach from Tong \emph{et al.}~\cite{tong2024}.
Instead of fixing a set of DD gates and optimizing over the DD sequence structure to find the best DD sequence~\cite{tong2024}, we optimize the rotational parameters entering in the chosen DD sequence to improve performance. This approach is similar to finding noise resilient quantum circuits through machine learning~\cite{cincio2021machine}. After introducing the theory behind such a learning approach to dynamical decoupling, we demonstrate on IBM Quantum hardware in two experiments that the optimized LDD sequences are better at suppressing errors than the CPMG, XY4 and UR6 sequences.  

\subsection{Theory}
We consider a quantum circuit described by the ideal quantum channels $\mathcal U_{j}$ applied in sequence,  
\begin{align}
\label{eq3:cleanQC}
\rho_{\text{ideal}}=\prod_{j}\mathcal U_{j}(\rho_{0}),
\end{align}
to an initial state $\rho_{0}=|\psi_{0}\rangle\langle\psi_{0}|$, which we assume is pure, to create a desired target state $\rho_{\text{ideal}}=|\psi_{\text{ideal}}\rangle\langle \psi_{\text{ideal}}|$. 
The quantum channels $\mathcal{U}_j$ may correspond to the ideal gates in a circuit.
Here, we concern ourselves with noise described by a collection of typically unknown quantum channels $\mathcal M_{j}$, in which we include potential mid-circuit measurements, that act between the unitary channels $\mathcal U_{j}$ when qubits are idling.
The noisy quantum circuit then takes the form $\prod_{j}\mathcal M_{j}\mathcal U_{j}$.
Each noise channel $\mathcal M_{j}$ can describe unitary and non-unitary errors.
To suppress noise induced by $\mathcal M_{j}$ we divide $\mathcal M_{j}$ into $N+1$ channels $\mathcal M_{j,k}$ so that $\prod_{k=1}^{N+1} \mathcal M_{j,k}=\mathcal M_{j}$, and insert a LDD sequence of the form shown in Fig.~\ref{fig2:dd_sequence}(c). 
The LDD sequence is described by parameterized unitary quantum channels $\mathcal R_{\vec{x}}(\cdot)=R_{\vec{x}}(\cdot)R_{\vec{x}}^{\dagger}$ that are applied between the $\mathcal M_{k}$'s where the LDD gates  $R_{\vec{x}}$ are  parameterized by $\vec{x}\in \mathbb R^{M}$. 
For simplicity, we assume that the LDD parameters $\vec{x}$ are the same for each LDD gate $R_{\vec{x}}$. 
If we include $N/2$ LDD gates between the noise channels $M_{j,k}$, followed by implementing $N/2$ times the corresponding inverse $R_{\vec{x}}^{\dagger}$ to insure that the total LDD sequence composes to the identity in the absence of noise, as depicted in Fig.~\ref{fig2:dd_sequence}, the noise channel $\mathcal M_{j}$ becomes 
\begin{align}
\label{eq4:LDDS}
\mathcal M_{j, \vec{x}}:=\mathcal M_{j,N+1}\prod_{k=\frac{N}{2}+1}^{N}\mathcal R_{\vec{x}}^{\dagger} \mathcal M_{j,k}\prod_{k=1}^{N/2}\mathcal  R_{\vec{x}} \mathcal M_{j,k}.
\end{align}
The state resulting from the noisy circuit with LDD is then given by 
\begin{align}
\label{eq5:ansatz}
\rho(\vec{x})=\prod_{j}\mathcal M_{j,\vec{x}}\mathcal U_{j}(\rho_{0}).
\end{align}
To optimize the parameters $\vec{x}$ we need a cost function $J(\vec{x})$ that is sensitive to the quality of the circuits in Eq.~\eqref{eq3:cleanQC}. 
For small circuits, a natural choice for a figure of merit or cost function is the fidelity error 
\begin{align}
\label{eq6:costfunction}
J(\vec{x})=1-\langle \psi_{\text{ideal}}|\rho(\vec{x})|\psi_{\text{ideal}}\rangle.
\end{align}
Here, $F=\langle \psi_{\text{ideal}}|\rho(\vec{x})|\psi_{\text{ideal}}\rangle$ is the fidelity with respect to the ideal state.
However, state tomography scales exponentially with system size.
For large circuits, a scalable cost function can be built in multiple ways. 
As done in Refs.~\cite{proctor2022, tong2024} we can invert the circuit with mirroring in which each $\smash{\mathcal{U}_j^\dagger}$ is applied in reverse such that $\rho_\text{ideal}=\rho_0$.
Alternatively, one can reduce the original quantum circuit to a Clifford circuit as done, for instance, in Refs.~\cite{czarnik2021, sack2024}.
Here, the single-qubit gates in a circuit (with the CNOT as the two-qubit gate) are replaced by Clifford gates such that the whole circuit is a Clifford gate.
In hardware where parameters are encoded in virtual-$Z$ gates~\cite{mckay2017}, this will preserve the structure and timing of the underlying pulses, thereby leaving most noise sources such as $T_1$, $T_2$ and cross-talk unchanged at the pulse-level.
We can then compute $\rho_\text{ideal}$ with efficient Clifford based simulators.

Solving the optimization problem
\begin{align}
\label{eq7:optimizationproblem}
\min_{\vec{x}\in\mathbb R^{M}}J(\vec{x}),
\end{align}
yields the optimal parameter values $\vec{x}^{*}$ of the LDD sequence that minimize $J(\vec{x})$. 
Since we do not know the noise processes described by $\mathcal M_{j}$, we minimize $J$ in an iterative, variational quantum algorithm type fashion by using quantum and classical computing resources in tandem \cite{cerezo2021,magann2021}. By measuring the output at the end of the quantum circuit we estimate $J$, while a classical search routine is employed to update the parameters. 

\subsection{Case studies on IBM hardware}
\label{sec:casestuddies}
We study the performance of LDD in two different experiments carried out on IBM Quantum hardware with Bell pairs; a valuable resource.
For instance, Bell pairs enable quantum gate teleportation~\cite{gottesman1999} and similar known states generated in a factory of resources enable circuit cutting~\cite{vazquez2024scaling}.
In the first experiment we suppress noise during mid-circuit measurements. 
In the second experiment we suppress noise resulting from an increasing circuit depth. 
In both cases we minimize fidelity loss due to noise by increasing the fidelity of preparing the Bell state $|\psi_{\text{ideal}}\rangle=|\Phi_{+}\rangle=\frac{1}{\sqrt{2}}(|00\rangle+|11\rangle)$, on two qubits $q_i$ and $q_j$ within a $n>2$ qubit system, starting from $\smash{\rho_{0}=|0\rangle\langle  0|^{\otimes n}}$.
To infer the value of $J(\vec{x})$ in Eq.~\eqref{eq6:costfunction} for the Bell state, it suffices to measure the expectation values of the $X_iX_j$, $Y_iY_j$ and $Z_iZ_j$ Pauli operators with respect to $\rho(\vec{x})$, rather than performing full state tomography~\cite{vogel1989determination,smithey1993measurement,leibfried2005creation,dunn1995experimental,hradil1997quantum,james2001measurement,smolin2012efficient}.
The cost function is thus 
\begin{align}
\label{eq8:explicit_form_of_J}
\begin{split}
J(\vec{x})&=1-\frac{1}{4} \langle \mathds{1}+X_iX_j-Y_iY_j+Z_iZ_j \rangle_{\rho(\vec{x})}.
\end{split}
\end{align}
The three correlators are each estimated with a quantum circuit  executed with 400 shots in each evaluation.
We employ parameterized decoupling operations given by arbitrary single-qubit rotations
\begin{align}
\label{eq9:singlequbitrotation}
R_{\vec{x}}=R(\theta, \phi, \lambda)=e^{-i\frac{\theta}{2}Z}e^{-i\frac{\phi}{2}Y}e^{-i\frac{\lambda}{2}Z},
\end{align}
where in both experiments we repeat $R(\theta,\phi,\lambda)$ and its inverse in total $N=4$ times, described by Eq.~\eqref{eq4:LDDS}.  
Here, we only consider three angles $\vec{x}=(\theta,\phi,\lambda)$ to parameterize the full LDD sequence.
All qubits with an LDD sequence thus share the same parameter values.
This limits the size of the search space.
On IBM Quantum hardware, these single-qubit rotations are implemented by three parameterized virtual-$Z$ rotations~\cite{mckay2017} and two $\sqrt{X}$ pulses.
Therefore, in LDD we optimize the angles in these virtual-$Z$ rotations.
We pick the Simultaneous Perturbation Stochastic Approximation (SPSA) gradient descent method~\cite{spall1992multivariate, spall1997accelerated} to solve the optimization problem in Eq.~\eqref{eq7:optimizationproblem} starting with $(\theta, \phi, \lambda)=(0,0,0)$. 
We allow SPSA a total of $100$ iterations.
At each iteration SPSA requires only two estimations of the objective function, regardless of the number of optimization parameters. The hyperparameter ``pertubation'' in SPSA is set to the default values and the hyperparameter ``learning rate'' is calibrated by the optimizer~\cite{kandala2017hardware}. 

Below we compare LDD with CPMG, XY4, and UR6 where DD sequences are inserted when qubits are idling. Throughout the two experiments we concatenate two LDD, CPMG and XY4 sequences, as shown in Fig.~\ref{fig2:dd_sequence}, and adjust the duration $\tau$ to match the idling times of the qubits.
Therefore, the LDD sequence is built from two $R_{\vec{x}}$ gates and two $R^\dagger_{\vec{x}}$ gates. At the hardware level, each $R_{\vec{x}}$ gate is implemented by three virtual-Z rotations and two $\sqrt{X}$ gates.
The $X$ and $Y$ gates are implemented by a single pulse with the same duration as a $\sqrt{X}$ gate.
An example of the transpiled DD sequences is shown in Appendix \ref{app:transpilation}.

\subsubsection{Suppressing noise during mid-circuit measurements}\label{sec:midcircuit}
A mid-circuit measurement (MCM) involves measuring qubits at intermediate stages within a quantum circuit.
MCMs have various applications including quantum error correction (QEC)~\cite{terhal2015}, quantum teleportation~\cite{bennett1993}, reducing the depth of a quantum circuit~\cite{baumer2023}, circuit cutting~\cite{vazquez2024scaling}, and analyzing complex quantum behaviour~\cite{koh2022experimental}. 
Unfortunately, MCMs may introduce noise on neighbouring qubits of the physical device~\cite{gupta2023probabilistic}. 

\begin{figure}[h]
     \centering     \includegraphics[width=0.7\linewidth, clip, trim=0 3 0 0]{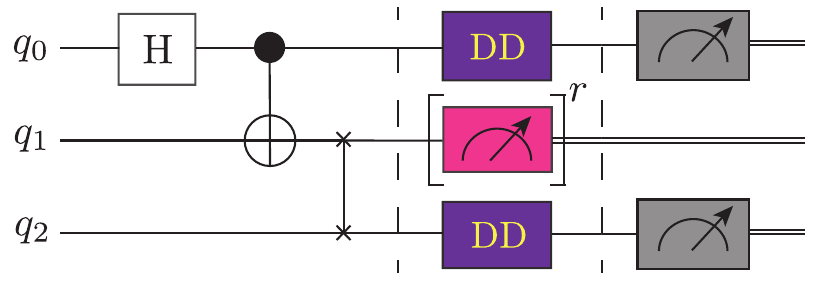}
     \caption{
     \textbf{Suppressing noise during MCMs through DD}. 
     A Bell state is prepared on qubit $q_0$ and $q_2$. The magenta block on qubit $q_1$ is a MCM that is repeated $r$ times. 
     The purple blocks on qubit $q_0$ and $q_2$ indicate DD sequences inserted during the MCMs to mitigate noise on $q_{0}$ and $q_{2}$.
     We vary $r$ and study the performance of different DD sequences on a $127$ qubit Eagle device (\textit{ibm\_kyiv}) and a $27$ qubit Falcon device (\textit{ibm\_hanoi}) reported in  Appendix ~\ref{app:mcm_hanoi}. 
     The corresponding experimental results are shown in Fig.~\ref{fig4:mcm_result} and Fig.~\ref{fig6:mcm_hanoi_deep_cairo}(a).}
    \label{fig3:mcm_circ}
\end{figure}

As depicted in Fig. \ref{fig3:mcm_circ}, we consider the task of preparing the Bell state $|\Phi_{+}\rangle$ between qubit $q_{0}$ and qubit $q_{2}$ while qubit $q_{1}$ is subject to $r$ repeated MCMs.
Here, varying $r\in\{1, ..., 15\}$ allows us to amplify the amount of noise introduced in the quantum circuit.  
To study the effect of DD during noisy measurements we first identify a measurement that introduces noise on its neighboring qubits, see  Appendix~\ref{app:noisy_mcm}, and thus map $(q_0,q_1,q_2)\to(q_{120}, q_{121}, q_{122})$. 

\begin{figure}[h]
     \centering     \includegraphics[width=0.85\columnwidth, clip, trim=0 0 0 0]{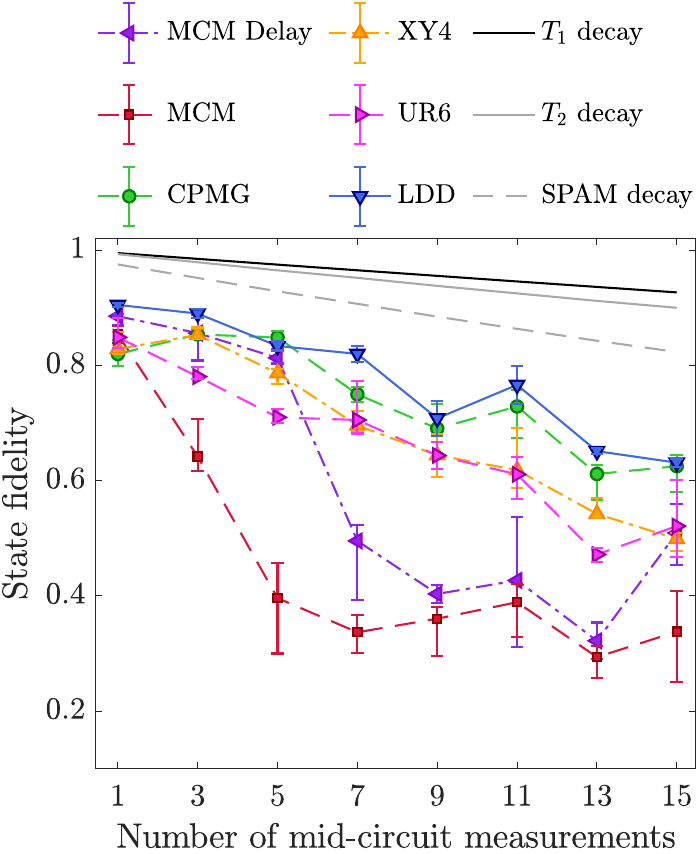}
     \caption{\textbf{DD noise suppression during MCMs.}
     Fidelity of a Bell state between two qubits as a function of the number of MCMs performed on another qubit, see Fig.~\ref{fig3:mcm_circ}. The dashed red curve shows the fidelity with MCMs but without DD and the dash-dotted purple curve shows the fidelity with a delay equivalent to MCMs.
     DD sequences are inserted when the qubits are idling. The dashed green curve corresponds to the CPMG sequence, the dash-dotted orange curve corresponds to the XY4 sequence, the dashed magenta curve corresponds to the UR6 sequence, and the solid blue curve corresponds to the optimized LDD sequence.
     For comparison, the solid black curve, the solid grey curve, and the dashed grey curve show the effect of $T_{1}$ decay, the $T_{2}$ decay, and the decay due to SPAM errors, respectively.  
     The error bars show the interquartile range.}
    \label{fig4:mcm_result}
\end{figure}

First, we compute $F$ without any DD, once with $r$ MCMs and once with a delay equivalent to $r$ MCMs. 
The experimental result are shown in Fig. \ref{fig4:mcm_result} (red and purple curves) where we report the median and the lower and upper quartiles of ten measurements.
In the presence of MCMs we observe a large drop in $F$ at $r=5$ from $\smash{0.813\substack{+0.016 \\ -0.01}}$ to $\smash{0.396 \substack{+0.061 \\ -0.096}}$.
This implies that a MCM on $q_{121}$ strongly impacts $q_{120}$ and $q_{122}$ aside from adding a $1244~{\rm ns}$ delay.
Crucially, the fidelity decrease due to the long delays and measurement-induced noise of the $r$ MCMs on $q_{121}$ can be mitigated by DD sequences inserted on neighbouring qubits $q_{120}$ and $q_{122}$ during the MCM measurement, see Fig.~\ref{fig3:mcm_circ}.
In Fig.~\ref{fig4:mcm_result} we compare the performance of LDD (blue), CPMG (green), XY4 (orange), and UR6 (magenta).
The LDD sequence yields the best performance, resulting in e.g., a fidelity of $0.631 \substack{+0.007 \\ -0.007}$ at $r=15$ while CPMG, XY4, and UR6 result in $0.625 \substack{+0.019 \\ -0.045}$, $0.499 \substack{+0.010 \\ -0.021}$, and $0.521 \substack{+0.08 \\ -0.054}$ respectively.
To evaluate the reliability of the LDD sequence we compute $F(\vec{x}^*)$ ten times after the learning and report the lower quartile, median, and upper quartile of these ten runs.
Due to queuing, these circuits were executed on the hardware two days after the optimal parameters $\vec{x}^*$ were learnt.
This indicates that the learned parameters $\vec{x}^*$ are stable in time.
We attribute the residual decay in part to $T_{1}$, $T_{2}$, and state preparation and measurement (SPAM) errors shown for comparison in Fig.~\ref{fig4:mcm_result} as a solid black, a solid grey, and a dashed grey curve, respectively. 
The $T_1$ curve is computed as $e^{-t/T_1}$ where $t$ is the idling time and $T_1=245~\mu{\rm s}$ is the average $T_1$ time of qubits $q_{120}$ and $q_{122}$ of \emph{ibm\_kyiv}. Similarly, the $T_{2}$  curve is computed as $e^{-t/T_2}$ where $T_2=175~\mu{\rm s}$ is the average $T_2$ time of qubits $q_120$ and $q_122$ of \emph{ibm\_kyiv}. 
The dashed grey line is $0.987e^{-t/T_2}e^{-t/T_1}$ where $0.987$ is the product of the readout assignment fidelity of qubits $q_{120}$ and $q_{122}$ as reported by the backend.

\subsubsection{Suppressing deep circuit noise}

\begin{figure*}[!t]
     \centering  \includegraphics[width=0.89\textwidth, clip, trim=0 3 0 0]{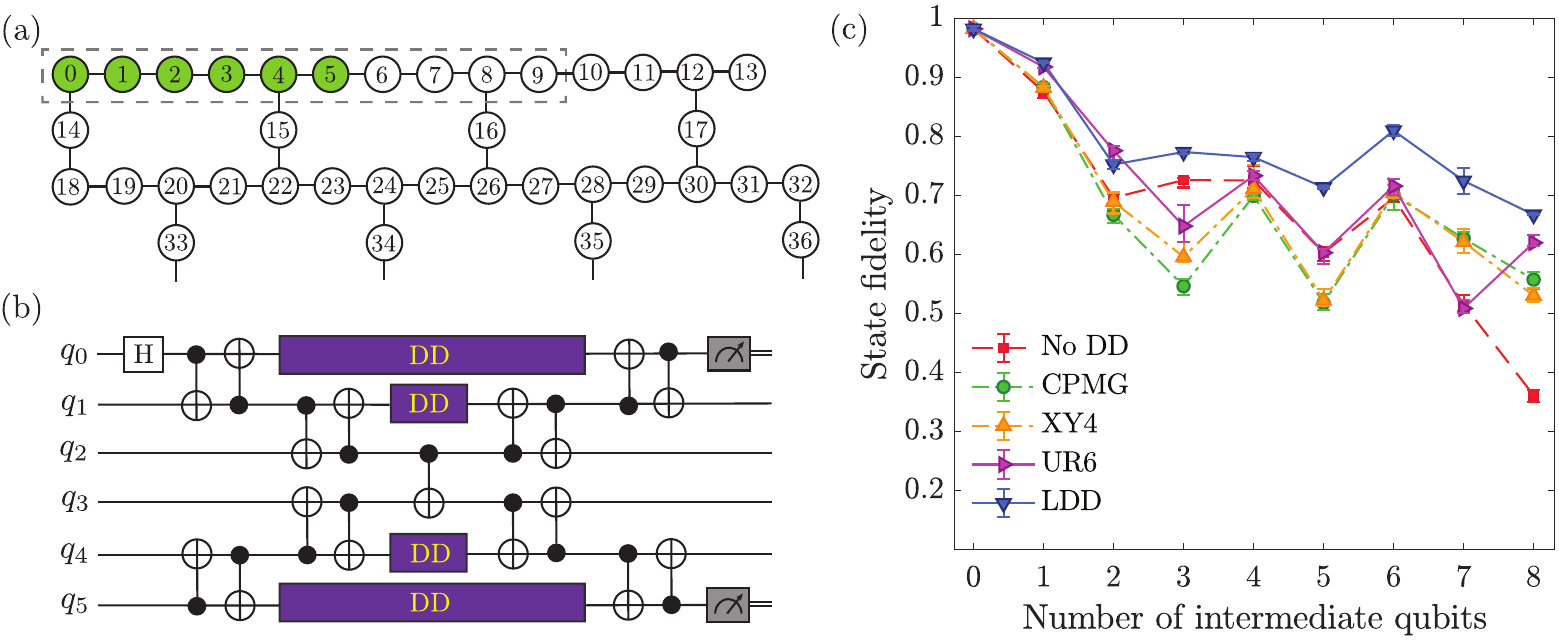}
     \caption{
     \textbf{Suppressing deep circuit noise through DD.} (a) A part of the qubit coupling graph of the $127$ qubit IBM Eagle device \textit{ibm\_strasbourg}.
     Qubits are shown as circles labeled by numbers. 
     The green circles highlight the qubits in the quantum circuit in (b). 
     The grey dashed box shows the chain of qubits $0, 1, 2, 3, 4, 5, 6, 7, 8$, and $9$ used in the experiment.    
    (b) Quantum circuit to prepare a Bell state between qubit $q_{0}$ (one edge of the chain) and qubit $q_{i}$ (here shown for $i=5$) using a ladder of CNOT gates. The insertion of DD sequences is shown as purple boxes. 
    (c) Fidelity of preparing a Bell state between qubit $q_{0}$ and $q_{i}$ in the chain in the coupling graph (a) as a function of the number of intermediate qubits. The dashed red curve shows the fidelity without DD, the dashed-dotted orange curve corresponds to the XY4 sequence, the dashed-dotted green curve corresponds to the CPMG sequence, the dashed magenta curve corresponds to the UR6 sequence, and the solid blue curve correspond to the optimized LDD sequence.
    The error bars show the interquartile range.}
    \label{fig5:deep_circ_analysis}
\end{figure*}

Next, we consider suppressing noise through LDD that is introduced due to increasing the depth of a quantum circuit. In particular, we consider the task of preparing the Bell state $|\Phi_{+}\rangle$ between two qubits located at the edges of a qubit chain with nearest neighbour interactions. 

The corresponding coupling graph of the IBM Quantum device is shown in Fig.~\ref{fig5:deep_circ_analysis}(a) where the considered qubit chain is highlighted in the dashed grey box. 
In Fig.~\ref{fig5:deep_circ_analysis}(b) we show the quantum circuit that prepares a Bell state between qubits $q_{0}$ and $q_{5}$ by bringing them into proximity with a ladder of SWAP gates.
Since the ancilla qubits are in their ground state $\ket{0}$, we can implement each SWAP gate with two CNOT gates instead of three.
Similar gate ladders with a single CNOT at each rung often occur in quantum simulation algorithms to create unitaries generated by  Pauli strings \cite{Tranter2018, Li2022}.
In Fig.~\ref{fig5:deep_circ_analysis}(c) we plot the fidelity $F$ for preparing the Bell state between the two edge qubits of the chain in Fig.~\ref{fig5:deep_circ_analysis}(a) as a function of the intermediate qubits (IQ) shown by the dashed red curve. 
We see a fidelity decrease from first for a chain consisting of $2$ qubits (i.e., $0$ IQ) to second for a chain consisting of $10$ qubits (i.e., $8$ IQ). 

To mitigate the Bell state fidelity decrease, as depicted in Fig.~\ref{fig5:deep_circ_analysis}(b), we insert DD sequences on idling qubits and compare their performance. The results shown in Fig.~\ref{fig5:deep_circ_analysis}(c) for the CPMG sequence (green), the XY4 sequence (orange), the UR6 sequence (magenta), and the LDD sequence (blue) suggest that the best performance is obtained for LDD.

In fact, since the fidelity obtained through inserting XY4 and CPMG can even be below the fidelity obtained without DD, we conclude that DD can increase the noise instead of suppressing it. 
This situation is avoided by LDD as the DD sequence is tailored to the device.

\section{Conclusion\label{sec:conclusion}}
Dynamical decoupling (DD) is a powerful noise suppression strategy that averages out detrimental processes by applying properly designed pulses to the system.
We introduced the framework of ``learning dynamical decoupling'' (LDD).
Instead of considering a DD sequence with fixed rotational gates, LDD optimizes directly on quantum hardware the rotational parameters in the DD gates. 
We compared the performance of such optimized DD sequences with the known DD sequences CPMG, XY4, and UR6 on IBM Quantum hardware. 
We found that LDD outperforms all three sequences in suppressing noise that occurs during mid-circuit measurements and noise that stems from increasing the depth of a quantum circuit. We believe that this is because the optimal angles in LDD are tailored to the noise of the device. We expect that this performance can be further improved by also optimizing over the DD gate spacings and the number of parameterized gates used.
Furthermore, the spacing between the DD gates and the way in which they are compiled into $R_Z$, $X$, and $\sqrt{X}$ gates may also impact performance~\cite{Vezvaee2024}.

The LDD sequences that we studied have by design a small number of single-qubit gates and a fixed number of rotational parameters. Appendix \ref{app:robust_param} shows that the optimized LDD parameters are robust against perturbations. While we believe that performance can be increased even further by adding more optimization parameters, the results shown in Fig.~\ref{fig5:deep_circ_analysis} for different system sizes (i.e., a different number of intermediate qubits) suggest that the number of LDD parameters can remain constant while achieving a similar performance when the system is scaled. As such, the classical optimization overhead does not need to increase when the system size increases. 
Therefore, the LDD approach considered here is scalable by design. 
Furthermore, inserting only a small number of single-qubit gates in LDD or DD on idling qubits to suppress noise is important for current quantum devices.
Idle times in quantum circuits typically occur when a subset of all qubits undergo two-qubit gates.
Therefore, the shorter the two-qubit gate duration is on a device, the more compact the DD sequence needs to be.
To illustrate this consider \emph{ibm\_torino} and \emph{ibm\_sherbrooke} which have a median two-qubit gate duration of $84~\mathrm{ns}$ and $533~\mathrm{ns}$, respectively~\footnote{The numbers are as reported from the backends. Furthermore, \emph{ibm\_torino} implements two-qubit CZ gates with tunable couplers while \emph{ibm\_sherbrooke} uses the cross-resonance interaction.}.
The duration of both the single-qubit $X$ and $\sqrt{X}$ gates are $32~\mathrm{ns}$ and $57~\mathrm{ns}$ for \emph{ibm\_torino} and \emph{ibm\_sherbrooke}, respectively.
As such, on \emph{ibm\_torino} we can insert up to two $X$ or $Y$ gates or one arbitrary single-qubit rotation during the median two-qubit gate duration.
By contrast, on \emph{ibm\_sherbrooke} these numbers become eight and four, respectively.
Consequently, as the two-qubit gate duration become comparable with the single-qubit gate duration, short DD sequences become more important.

In summary, DD is crucial to suppressing errors in noisy hardware.
As DD sequences improve -- becoming more tailored to the hardware -- so do the hardware results.
This motivates the strong interest in DD.
Future work may include optimizing the spacing of the DD pulses in LDD.
Furthermore, one could explore how to protect circuit cutting resources, consumed in teleportation circuits as demonstrated in Ref.~\cite{vazquez2024scaling}.
Indeed, these resources are less costly to generate simultaneously.
However, this has the draw-back that they idle until they are consumed. 
Finally, we optimized virtual-$Z$ rotations that sandwich $\sqrt{X}$ gates.
Future work may thus elect to directly optimize the pulses that implement the DD sequence, e.g., similar to pulse-level variational quantum algorithms~\cite{magann2021, egger2024pulse}.\\

\vspace{-0.75cm}
\begin{acknowledgments}
\vspace{-0.25cm}
C. A. acknowledges support from the National Science Foundation (Grant No. 2231328). A. R. and C. A. acknowledge support from Knowledge Enterprise at Arizona State University. 
D. J. E. acknowledges funding within the HPQC project by the Austrian Research Promotion Agency (FFG, project number 897481) supported by the European Union – NextGenerationEU.
We acknowledge the use of IBM Quantum services for this work. 
The views expressed are those of the authors, and do not reflect the official policy or position of IBM or IBM Quantum.
\end{acknowledgments}

\bibliography{citation}

\onecolumngrid
\appendix

\section{ Identifying qubits with the most MCM-induced noise}\label{app:noisy_mcm}

\begin{figure}[!ht]
     \centering    \includegraphics[width=0.7\columnwidth]{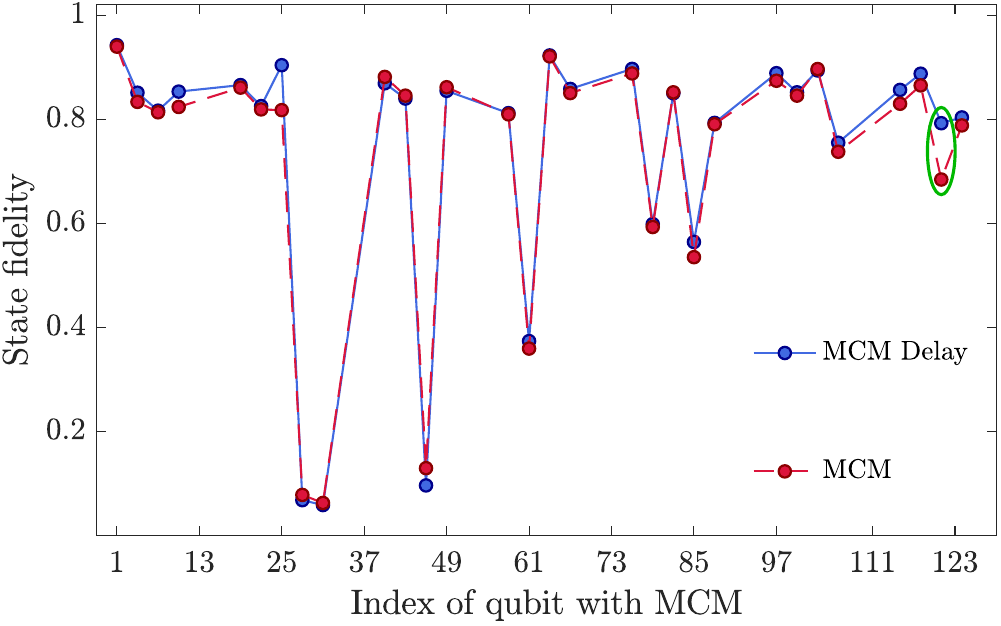}    
     \caption{\textbf{Noisy MCM identification experiment.}
     Bell state fidelity between qubits $q_{0}$ and $q_{2}$ as a function of the index of qubit $q_{1}$ in Fig.~\ref{fig3:mcm_circ} of the manuscript.
    The green ellipse shows qubit $121$ for which the state fidelity with a MCM (red curve) is significantly lower than when there is a delay equivalent to a MCM (blue curve).}
     \label{fig7:mcm_effect}
\end{figure}

The amount of noise an MCM introduces depends on the qubit on which the MCM is performed. To demonstrate that LDD can suppress noise introduced through MCM's, we identify qubits in the device where MCM's have a large effect on the fidelity of a Bell state. This is achieved by identifying qubits on the 127 qubit IBM device (\emph{ibm\_kyiv}) for which repeated MCM's significantly introduce more noise than letting the qubit idle for the corresponding amount of time. These qubits were found by preparing a  Bell state between two qubits, followed by inserting a MCM ($r=1$) or the delay equivalent to a MCM ($r=0$) on another qubit, as shown in Fig.~\ref{fig3:mcm_circ}. 
In the experiment, we choose the indices of the three qubit such that their corresponding physical qubits are adjacent to each other. 
This ensures that the noise in the circuit is introduced by the MCM rather than by the circuit transpilation. 
The Bell state fidelity is shown as a function of the qubit index on which the MCM or delay are applied in Fig.~\ref{fig7:mcm_effect}. 
The indices of the qubits with MCM that we consider in \textit{ibm\_kyiv} are $1, 4, 7, 10, 19, 22, 25, 28, 31, 40, 43, 46, 49, 58, 61, 64, 67, 76, 79, 82, 85, 88, 97, 100, 103, 106, 115, 118, 121$ and $124$. These choices ensure that the total three-qubit system forms a chain with nearest neighbour interactions in the coupling map of the device to avoid any additional SWAP gates that may introduce more noise.    

The red curve in Fig.~\ref{fig7:mcm_effect} shows the state fidelity in the presence of a MCM while the blue curve corresponds to the same circuit without an MCM, i.e., the MCM is replaced by a delay with the same duration. 
The qubit candidate (index $121$) chosen for the MCM experiments whose results are shown in Fig. \ref{fig4:mcm_result} in the main text is highlighted by a green ellipse.

\section{Additional data\label{app:mcm_hanoi}}

\subsection{MCM noise}
\begin{figure}[h]
     \centering     \includegraphics[width=0.85\columnwidth, clip, trim=0 0 0 0]{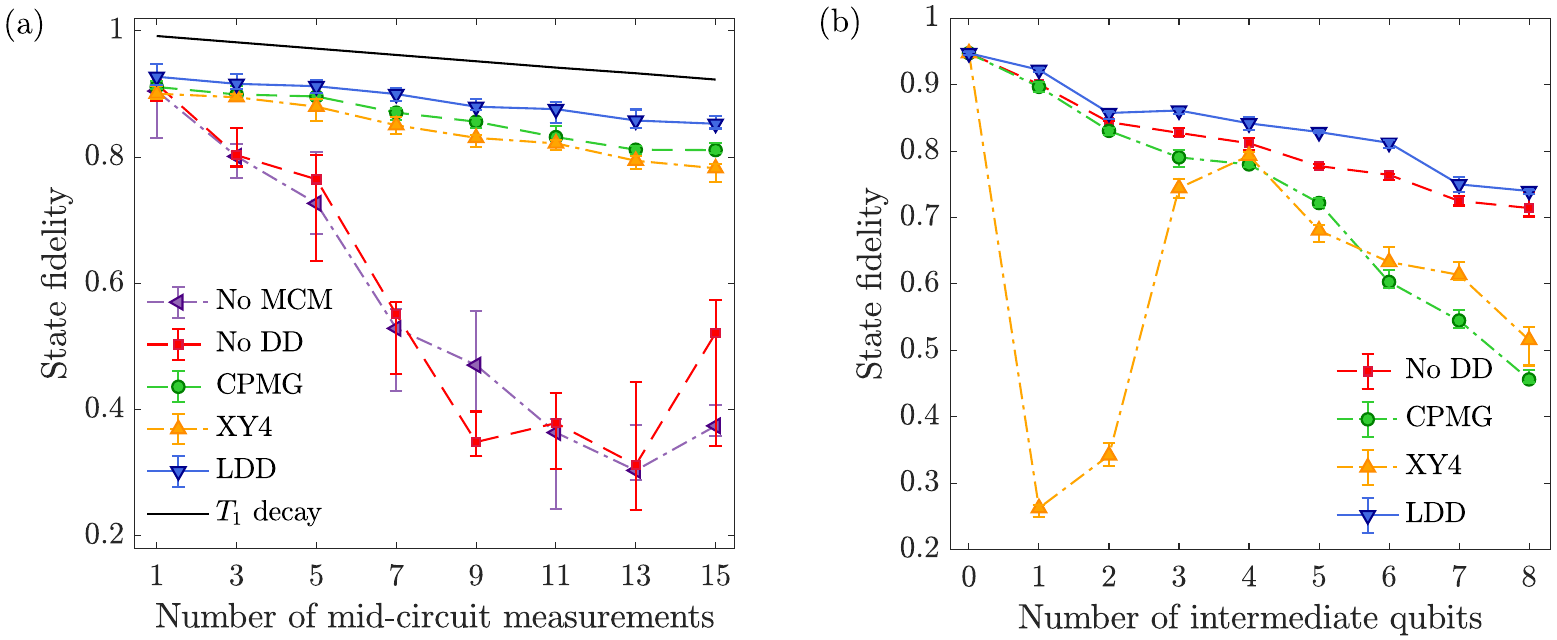}
     \caption{\textbf{DD noise suppression during MCMs and deep circuit noise suppression using \textit{ibm\_hanoi} and \textit{ibm\_cairo} respectively.}
     (a) Fidelity as a function of the number of MCMs performed. The dashed red curve shows the fidelity with MCMs but without DD and the dash-dotted purple curve shows the fidelity with a delay equivalent to MCMs. The dashed green curve corresponds to the CPMG sequence, the dash-dotted orange curve corresponds to the XY4 sequence and the solid blue curve corresponds to the optimized LDD sequence. The optimal LDD parameters are shown in Table~\ref{tab1:mcm}.
     For comparison, the solid black curve shows the $T_{1}$ decay. 
     The error bars show the interquartile range. (b)
     Fidelity of preparing a Bell state between two qubits in the chain in the coupling graph as a function of the number of intermediate qubits.  The dashed red curve shows the fidelity without DD, the dashed-dotted orange curve corresponds to the XY4 sequence, the dashed-dotted green curve corresponds to the CPMG sequence, and the solid blue curve correspond to the optimized LDD sequence. The optimal LDD parameters are shown in Table \ref{tab2:deep_circ}.
     The error bars show the interquartile range.}
    \label{fig6:mcm_hanoi_deep_cairo}
\end{figure}

In addition to the data presented in the main text obtained from the 127 qubit Eagle device \emph{ibm\_kyiv}, we also gathered data from the 27 qubit Falcon device \emph{ibm\_hanoi} for LDD, CPMG, and XY4.
Here, we prepared Bell state between qubit $q_0$ and $q_2$, inserted MCM on qubit $q_1$.
The results for the effect of a MCM on qubit $q_0$ and $q_2$ are shown in Fig.~\ref{fig6:mcm_hanoi_deep_cairo}(a).
We report the median and the lower and upper quartiles of ten measurements.
In the presence of MCMs (red curve), we observe a large drop in $F$ from $\smash{0.916\substack{+0.004 \\ -0.026}}$ at $r=1$ to ${0.521 \substack{+0.052 \\ -0.178}}$ at $r=15$. Without MCMs (purple curve) $F$ drops from ${0.905 \substack{+0.015\\ -0.075}}$ at $r=1$ to ${0.374 \substack{+0.033 \\ -0.016}}$ at $r=15$. 
This implies that on \emph{ibm\_hanoi} a MCM on a single qubit does not have a large impact on the adjacent qubits aside from adding a $820~{\rm ns}$ delay.
These qubits were not selected using the procedure used to identify noisy MCMs on \emph{ibm\_kyiv}.

In Fig.~\ref{fig6:mcm_hanoi_deep_cairo}(a), we compare the performance of LDD (blue), CPMG (green), and XY4 (orange).
The LDD sequence, whose corresponding optimal parameters are shown in Table \ref{tab1:mcm}, yields the best performance, resulting in e.g., a fidelity of ${0.855 \substack{+0.012 \\ -0.007}}$ at $r=15$ while CPMG and XY4 result in ${0.811 \substack{+0.012 \\ -0.005}}$ and ${0.782 \substack{+0.007 \\ -0.021}}$, respectively.
We attribute the residual decay to $T_{1}$ shown for comparison in Fig.~\ref{fig6:mcm_hanoi_deep_cairo}(a) as a solid black curve, where $T_1=167~\mu{\rm s}$ is the average $T_1$ time of qubits $q_0$ and $q_2$ of \emph{ibm\_hanoi}.

\begin{table}[!h]
\begin{ruledtabular}
\begin{tabular}{c l c c c} 
{\text{\#} MCMs} & {Idle time} & \multicolumn{3}{c}{\text{Optimal LDD parameters}}\\
 & {($\mu s$)} & $\theta^{*}$ & $\phi^{*}$ & $\lambda^{*}$ \\
\hline
1 & 0.82  &  0.37$\pi$ &  0.22$\pi$ &  0.05$\pi$  \\
3 & 2.45  & -0.32$\pi$ & -0.35$\pi$ &  0.13$\pi$ \\
5 & 4.09  & -0.66$\pi$ &  0.74$\pi$ & -0.66$\pi$ \\
7 & 5.72  & -0.67$\pi$ &  0.40$\pi$  & -0.54$\pi$ \\
9 & 7.35  & -0.28$\pi$ & -0.53$\pi$ & -0.29$\pi$ \\
11& 8.99  & -1.13$\pi$ &  1.49$\pi$ & -1.34$\pi$  \\
13& 10.62 &  0.95$\pi$ & -1.07$\pi$ & -0.68$\pi$ \\
15& 12.26 & -0.99$\pi$ & -0.79$\pi$ &  0.50$\pi$ \\
\end{tabular}
\end{ruledtabular}
\caption{\textbf{Optimal LDD parameters to suppress noise during MCMs.} 
The angles in the LDD gate given in Eq.~\eqref{eq9:singlequbitrotation} are optimized to  find the optimal rotational (Euler) angles $\Vec{x}^{*}=(\theta^{*},\phi^{*},\lambda^{*})$. 
The optimal angles are shown for a different number of MCMs applied on $q_{1}$ that correspond to the idling times of qubit $q_{0}$ and $q_{2}$. 
For reference, the Euler angles for the Pauli $X$, and Pauli $Y$ gates used in the XY4 sequence are $(\pi, \pi, 0)$ and $(\pi, 0, 0)$, respectively.}
\label{tab1:mcm}
\end{table}

\subsection{Deep circuit noise}
In addition to the data presented in the main text on the 127 qubit Eagle device \emph{ibm\_strasbourg}, we also gathered data on the 27 qubit Falcon device \emph{ibm\_cairo} for LDD, CPMG, and XY4, see Fig. \ref{fig6:mcm_hanoi_deep_cairo}(b).
We plot the fidelity $F$ for preparing the Bell state between the two edge qubits of the qubit chain in \textit{ibm\_cairo} as a function of the intermediate qubits (IQ) shown by the dashed red curve. 
We see a fidelity decrease from ${0.948 \substack{+0.002 \\ -0.002}}$ for a chain consisting of $2$ qubits (i.e., $0$ IQ) to ${0.720 \substack{+0.024 \\ -0.013}}$ for a chain consisting of $10$ qubits (i.e., $8$ IQ). 

To mitigate the Bell state fidelity decrease, we insert DD sequences on idling qubits and compare their performance. The results shown in Fig.~\ref{fig6:mcm_hanoi_deep_cairo}(b) for the CPMG sequence (green), the XY4 sequence (orange), and the LDD sequence (blue) suggest that the best performance is obtained for LDD where the corresponding optimal parameters are given in Table~\ref{tab2:deep_circ}.

\begin{table}[h]
\begin{ruledtabular}
\begin{tabular}{ c  r  r  r }
{\text{\#} IQs} & \multicolumn{3}{c}{\text{Optimal LDD parameters}}\\
 & $\theta^{*}$ & $\phi^{*}$ & $\lambda^{*}$ \\
\hline
1 & -0.41$\pi$ &  0.47$\pi$ & -0.28$\pi$ \\
2 & -0.03$\pi$ &  0.72$\pi$ &  0.03$\pi$ \\
3 & -0.02$\pi$ & -0.45$\pi$ &  0.02$\pi$ \\
4 & -0.07$\pi$ & -0.47$\pi$ & -0.25$\pi$ \\
5 & -0.03$\pi$ & -0.48$\pi$ & -0.17$\pi$ \\
6 & -0.02$\pi$ &  0.24$\pi$ &  1.02$\pi$ \\
7 &  1.94$\pi$ &  0.85$\pi$ & -0.42$\pi$ \\
8 & -0.01$\pi$ &  1.85$\pi$ & -0.33$\pi$ \\
\end{tabular}
\end{ruledtabular}
\caption{\textbf{Optimal LDD parameters for suppressing deep circuit noise.} The angles in the LDD gate given in Eq.~\eqref{eq9:singlequbitrotation} are optimized to  find the optimal rotational (Euler) angles $(\theta^{*},\phi^{*},\lambda^{*})$. The optimal $\Vec{x}^*=(\theta^{*}, \phi^{*}, \lambda^{*})$ are shown for different number of intermediate qubits (IQ).}
\label{tab2:deep_circ}
\end{table}

\section{Robustness of the optimal angles to perturbations}\label{app:robust_param}
To study the robustness of the optimized angles we perform additional MCM experiments on \emph{ibm\_kyiv}.  
\begin{figure}[!ht]
     \centering    \includegraphics[width=0.48\columnwidth]{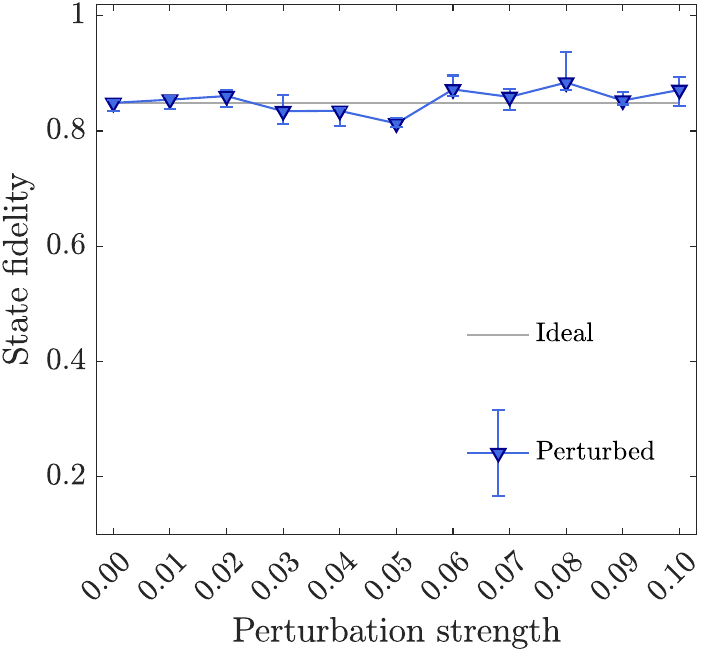}    
     \caption{\textbf{Robustness of the optimal angles to pertubations.} The Bell state fidelity between qubits $q_{120}$ and $q_{122}$ as a function of the strength of the perturbation $\epsilon$, see Eq. \eqref{eq:perturbed} and Eq. \eqref{eq:perturbedLDD}. The fidelity corresponding to the optimal angles ($\epsilon=0$) is shown as a grey line.} 
     \label{fig8:robust_param}
\end{figure}
We first optimize the angles $\theta,\phi,\lambda$ in a LDD sequence for $r=1$ as in Fig.~\ref{fig3:mcm_circ}. 
Next, we investigate the robustness by perturbing the angles in $R_{\vec{x}}=R(\theta, \phi, \lambda)$ and reporting the effect on the fidelity.

In particular,  we uniformly sample three parameters $\Delta\theta$, $\Delta\phi$, and $\Delta\lambda$ from the interval $[-2\pi,2\pi]$.
We normalize the resulting vector $\vec{\delta} = [\Delta\theta, \Delta\phi, \Delta\lambda]$ to have a norm of $\epsilon$ which controls the magnitude of the perturbation, i.e.,  
\begin{align}
\label{eq:perturbed}
\vec{\Delta} = \epsilon \ \frac{\vec{\delta}}{||\vec{\delta}||} = [\Delta\theta^{'}, \Delta\phi^{'}, \Delta\lambda^{'}].
\end{align}
We perform experiments for the perturbed LDD gates, 
\begin{align}
\label{eq:perturbedLDD}
R_{\vec{x}+\vec{\Delta}}=R(\theta+\Delta\theta^{'}, \phi+\Delta\phi^{'}, \lambda+\Delta\lambda^{'}),
\end{align}
as a function of the perturbation strength $\epsilon$. 
The results are shown in Fig.~\ref{fig8:robust_param} where each data point corresponds the median of $10$ samples and the error bars show the upper and lower quartile range. 
For comparison we show the fidelity found without perturbing the optimized angles as a grey line.

\newpage
\section{DD transpiled circuits\label{app:transpilation}}
\begin{figure}[!h]
     \centering    \includegraphics[width=0.8\columnwidth]{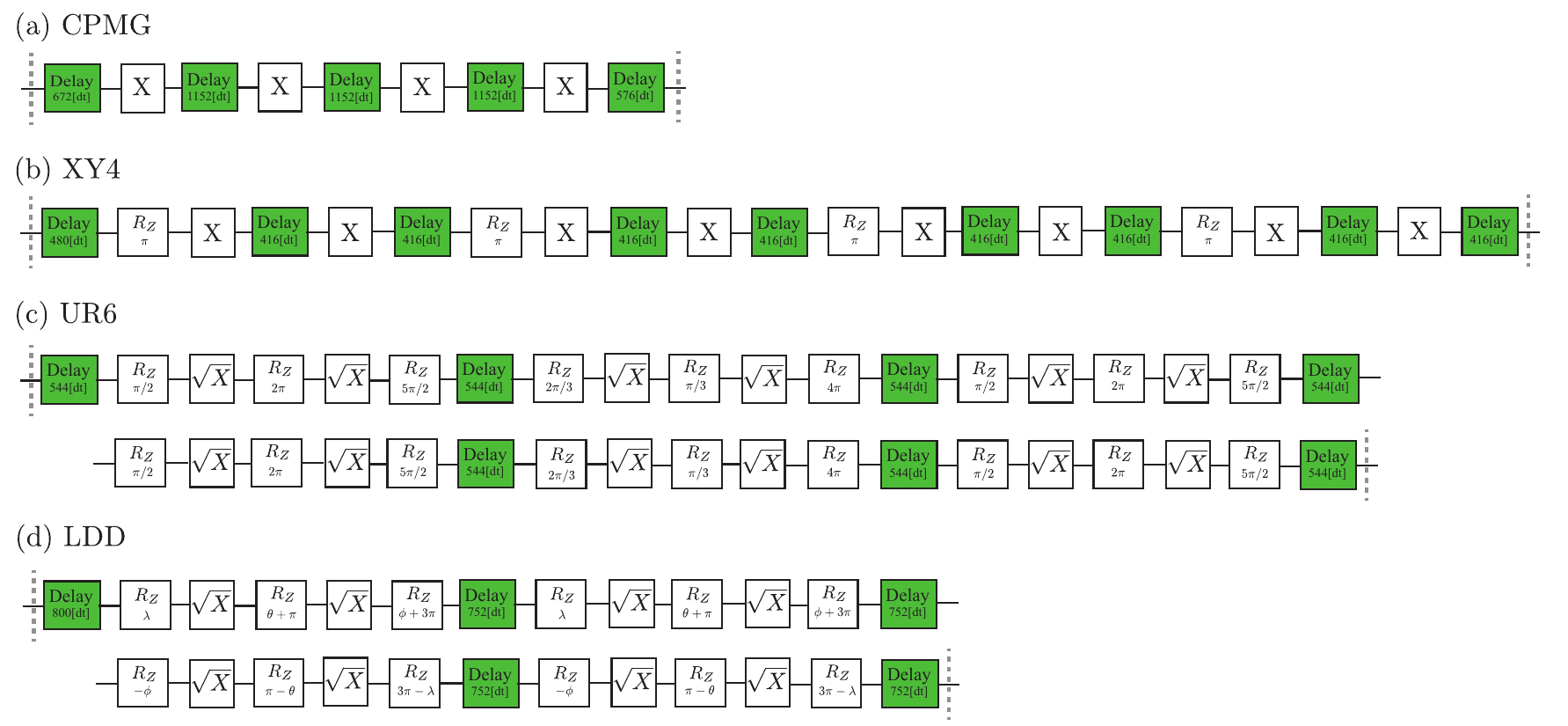}    
     \caption{\textbf{Transpiled DD sequences.} 
     (a) CPMG, (b) XY4, (c) UR6, and (d) LDD inserted in an idle time of the same length as a mid-circuit measurement of duration $1244~{\rm ns}$, i.e., $5600 [dt]$ where $dt=0.22 \ n$s. The hardware native gates are $X, \sqrt{X}, \ \text{and} \ R_Z(\theta )$. 
     }
     \label{fig:transpiled_circuit}
\end{figure}

\begin{figure}[!h]
     \centering    \includegraphics[width=0.7\columnwidth]{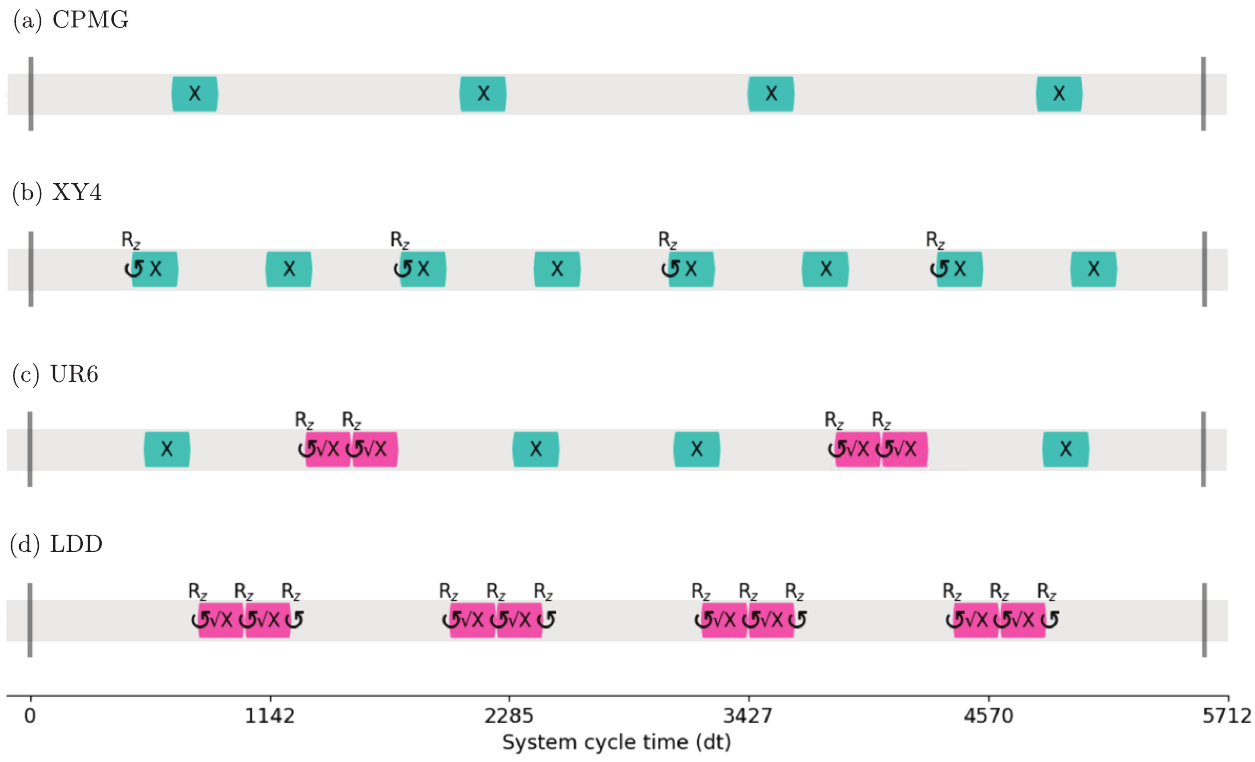}    
     \caption{\textbf{Timeline drawing of the transpiled DD sequences.} The (a) CPMG, (b) XY4, (c) UR6, and (d) LDD sequences are the same as those in Fig.~\ref{fig:transpiled_circuit} and plotted as a function of time in units of $dt$ on the $x$-axis.
     Here, the width of the $X$ and $\sqrt{X}$ gates, shown in green and red, respectively, matches their duration.
     The $R_Z(\theta)$ gates, shown as curved arrows, are virtual and thus have a duration of zero.
     }
     \label{fig:timeline_drawing}
\end{figure}

The circuits in this work are transpiled to the qubit coupling map and the hardware native gates of the quantum device. 
To insert a DD sequence in a circuit we first schedule the circuit.
I.e., we attach to the hardware native circuit instructions their corresponding duration and schedule them as late as possible.
Next, idling times on the qubits that are long enough are replaced by delays and DD gates.
The duration of the delays are determined by the total idle time and the DD sequence. 
In this appendix we exemplify the gate sequences resulting from this transpilation process for a delay of $1244~{\rm ns}$ which corresponds to the duration of a mid-circuit measurement on \textit{ibm\_kyiv}, see also Sec.~\ref{sec:midcircuit}.
We implement CPMG, XY4, UR6, and LDD to fit this idling time. 
Fig. \ref{fig:transpiled_circuit} shows the transpiled circuits and Fig. \ref{fig:timeline_drawing} shows their timeline diagram with time measured in units of system cycle time.

\end{document}